# A Numerical Study on the Effects of Heterogeneity, Anisotropy, and Station Coverage on the Compensated Linear Vector Dipole Component of Deep Earthquake Moment Tensors

Jiaxuan Li[1*], Yingcai Zheng[1], and Xinding Fang[1,2]

[1] Department of Earth and Atmospheric Sciences, University of Houston, Houston, TX, 77204-5007, USA

[2] Department of Ocean Science and Engineering, Southern University of Science and Technology of China, No 1088, Xueyuan Rd., Nanshan District, Shenzhen, Guangdong 518055, China

*Correspondence to: jli74@uh.edu

# 1 Abstract

The moment tensors of a large portion of deep earthquakes show apparent non-double-couple (non-DC) components. Previously, the observed apparent non-DC values in deep earthquakes have been attributed to different mechanisms such as complex source processes or complicated source medium structures. In this paper, we focused on evaluating the second mechanism. We investigated the effect of slab heterogeneity, supra-slab anisotropic structure, intra-slab weakly anisotropic structure (e.g., the purported existence of the metastable olivine wedge), and non-uniform station coverage, on the non-DC radiation patterns of deep earthquakes using our 3-

dimensional elastic finite-difference modeling and full-waveform inversion of moment tensors. We found that these investigated issues cannot cause the observed non-double-couple radiation patterns and the in-situ structure with strong S-wave anisotropy near to the earthquake focus is the simplest way to account for the apparent non-DC components in the radiation patterns of deep earthquakes.

## 2 Introduction

Seismic moment tensors have been routinely determined either locally or globally by different catalogs (Dziewonski *et al.*, 1981; Kubo *et al.*, 2002; Ekström *et al.*, 2012). The simple linear relationship between the amplitude of free oscillation and the seismic moment tensor established by Gilbert (1971), and the broad deployment of broadband seismometers contribute to this routine process. However, routine determination of moment tensors is based on the assumption that the excitation kernel functions (based on 1d model) are relatively exact and could correctly represent the earth structure (Dziewonski *et al.*, 1981). Although the effect of major structures on wave propagation is well understood, consideration of the effect of small-scale near-source structure has been lacking.

An earthquake moment tensor can be decomposed as a linear summation of three parts: a DC, a volumetric (or isotropic), and a compensated linear vector dipole (CLVD) component. The non-DC component consists of the latter two parts. The CLVD component in the moment tensor of deep earthquakes is commonly observed. Since Knopoff and Randall (1970) and Randall and Knopoff (1970) first proposed that there are significant CLVD components in moment tensors inverted using body waves of several deep earthquakes, several observations have confirmed the existence of a CLVD component in deep earthquakes (Kuge and Kawakatsu, 1993; Henry *et al.*,

2002). However, there are still doubts about what exactly causes the CLVD component. Since a slip on a planar fault in an isotropic homogeneous elastic medium generates the DC seismic radiation, the observed non-DC radiation patterns have been attributed to possible mechanisms including complex fault geometries (Frohlich, 1990; Julian *et al.*, 1998) and complex source structures. The complex source structure could either be unknown heterogeneous structure in the source region (Woodhouse, 1981; Johnston and Langston, 1984; Tada and Shimazaki, 1994) because most deep earthquakes are embedded in heterogeneous subducting slabs, or in-situ anisotropic structures (Kawasaki and Tanimoto, 1981; Vavrycuk, 2004; Li *et al.*, 2018).

The main purpose of this numerical study is to analyze the effect of near-source medium properties such as heterogeneity and anisotropy on the CLVD component in the inverted moment tensors of deep earthquakes. In our modeling, we assume all earthquakes have simple pure shear dislocations on a planar fault in either isotropic or anisotropic media. We will model synthetic seismograms by applying a 3-D full-elastic staggered-grid finite-difference (FD) modeling (Fang *et al.*, 2017) using 24 different source fault geometries in five different models.

We will show that the heterogeneity near the source only has a minor effect on the CLVD component of inverted moment tensors using a reasonable station coverage. The inverted CLVD component thus represents other source-region properties such as the in-situ anisotropy. Since the metastable olivine wedge (MOW) has been prevalently invoked to explain for the mechanism of deep-focus deep earthquakes (Bina and Wood, 1987; Jiang *et al.*, 2008; Green *et al.*, 2010; Kawakatsu and Yoshioka, 2011) the effect of MOW is tested here. It is shown that MOW alone could not produce large CLVD components in inverted moment tensors.

In addition, we will show that the station coverage does play an important role in moment tensor determination (Satake, 1985). A poor station coverage could generate artifacts in the

CLVD components when the effects of heterogeneity are neglected in kernel functions. However, if we consider the effect of heterogeneous structure in the kernel functions, we could retrieve more precise CLVD values even with a poorer station coverage.

# 3 Methods:

## 3.1 Model Setup

We performed the FD forward modeling using a Ricker wavelet with a central frequency of 0.1Hz as source time function. Because our purpose is to investigate the near-source effects, we place 441 receivers, evenly distributed on a sphere with a radius of 500km centered at the source location (Figure 1a). We will model seismic wavefields in different models to study the structural effect on the moment tensor inversion. For Model-0, the background medium is isotropic homogeneous with a P-wave velocity 8km/s and S-wave velocity 4.5km/s. For the first three cases investigated in the following section, we used the waveforms recorded in Model-0 as our Green's function in the inversion, which is similar to the kernel function used in routine inversion. Note that, this Green's function only contains information of background medium properties. No source-region elastic heterogeneous and/or anisotropic structure information are included in the Green's function.

We also did FD modeling for five other models which introduced heterogeneous and/or anisotropic structures (Figure 2). For anisotropic structures, we focused on a particular type of anisotropy with tilted transversely isotropic (TTI) symmetry (Li *et al.*, 2018). One such example is a rock with fine layering fabric. The TTI anisotropy is characterized by two angles, $\theta, \varphi$, for the orientation of the symmetry axis plus five constants denoted by the *P*- and *S*-wave velocities propagating along the symmetry axis and three Thomsen parameters, $\varepsilon, \gamma, and\ \delta$ (Thomsen,

1986), with ε measuring the P-wave anisotropy, γ the S-wave anisotropy, and δ affecting both P and S waves. These five models are:

(i) Model-1 (Figure 2a) embeds/adds a 600km × 600km × 100km (thickness) velocity-heterogeneous slab in Model-0, with strike 180° and dip 40°. The $V_P$ and $V_S$ in the coldest part of the slab is 5% higher than the background velocities (Figure 1b). The source is located in the center of the slab, 20km away from the upper interface of the slab.

(ii) Model-2 (Figure 2b) has no velocity-heterogeneous slab but adds a 30km × 30km × 30km highly anisotropic patch around the source region. The anisotropic patch is a TTI medium (Li *et al.*, 2018). The TTI symmetry axis is perpendicular to the slab interface in Model-1. The 5 independent elastic parameters of this TTI medium are described by the $V_P$ and $V_S$ along the TTI symmetry axis and three Thomsen parameters ε, γ, δ. Here, $V_P = 8.4 km/s$, $V_S$=4.7 km/s, $\varepsilon = 20\%, \gamma = 20\%, \delta = 0\%$. The source is at the same location as the one in Model-1 and is located in the center of the anisotropic patch.

(iii) Model-3 (Figure 2c) adds both velocity-heterogeneous slab and anisotropic patch described in Model-1 and Model-2. The source has the same location as the one in Model-1.

(iv) Model-4 (Figure 2d) adds the same velocity-heterogeneous slab described in Model-1 but has an supra-slab weakly anisotropic patch atop the slab. This anisotropic patch has a dimension of 150km × 150km × 20km (thickness). The orientation of the TTI symmetry axis is the same as the one in Model-2 and Model-3. $V_P$ and $V_S$ in the anisotropic medium along the TTI symmetry axis are the same as the background $P$ and $S$ velocities, respectively. The Thomsen parameters are $\epsilon = 0\%, \gamma = 6\%, \delta = 0\%$. The source has the same location as the one in Model-1.

(v) Model-5 (Figure 2e) adds the same velocity-heterogeneous slab described in Model-1. In order to mimic the possible existence of metastable olivine wedge (MOW) (Furumura *et al.*, 2016) in the earthquake source region, we add a weakly anisotropic wedge. The MOW is also a TTI medium. Its TTI symmetry axis orientation is parallel to the slab interface. The $V_P$ and $V_S$ along the TTI symmetry axis inside the MOW are about 5% slower than their background values. The Thomsen parameters are $\epsilon = 0\%, \gamma = 5\%, \delta = 0\%$. We show a cross-sectional profile of the MOW (Figure 2f), the distances of its three tips A, B, and C from the upper slab interface are 20km, 0km, and 40km. The gray lines in Figure 2f represent the layered anisotropic structure inside the MOW. The source has the same location as the one in Model-1.

## 3.2 Moment Tensor Inversion Methods

We randomly generated 24 shear dislocation sources with different fault geometries. We then compute the moment tensors (Aki and Richards, 2002) using the elastic tensor and fault geometry in the following equation (1). In this equation, $i, j, k, l = 1,2,3$ represent the 3 spatial directions, $u$ is the magnitude of the slip of the dislocation faulting, $S$ is the area of the slippage, $c_{ijkl}$ is the elastic tensor in the source region, $n_k$ is the fault unit normal, and $v_l$ is the slip unit vector.

$$m_{ij} = usc_{ijkl}n_k v_l, \quad i,j,k,l = 1,2,3 \ . \tag{1}$$

These synthetic moment tensors were used in our FD modeling in each Model-$J$ (e.g. Model-$J$, $J = 1,2,3,4,5$) to produce synthetic waveforms, denoted by $u_k^J(\mathbf{r}, t)$, which represents the $k$-th component of the displacement recorded at receiver $\mathbf{r}$. The relationship between the displacement field, the moment tensor and Green's function $G$ (or the kernel function) can be expressed as (Aki and Richards, 2002).

$$u_k^J(\mathbf{r},t) = G_{kp,q}^J(\mathbf{r},t) m_{pq} \ . \tag{2}$$

The Green's function $G_{kp}^J(\mathbf{r}, t)$ represents the $k$-th component of the displacement at $\mathbf{r}$ due to a single force along the $p$-th direction at the source position in Model-$J$. The subscript in equation (2), $q$, represents the spatial derivative of the Green's function along $q$-th direction for the source location.

After we obtained synthetic waveforms, a least-squares method was applied to invert for the moment tensors using the Green's function obtained in the homogeneous isotropic Model-0. We minimize the objective function $L$, to determine the moment tensor $m_{pq}$ and the time shift $\tau_\mathbf{r}$ at receiver $\mathbf{r}$

$$L\left(m_{pq}^J, \tau_\mathbf{r}^J\right) = \sum_\mathbf{r} \sum_{k=1}^{3} \int \left[u_k^J\left(\mathbf{r},t+\tau_\mathbf{r}^J\right) - m_{pq}^J G_{kp,q}^0\left(\mathbf{r},t\right)\right]^2 dt \ . \tag{3}$$

The objective function represents the difference between the observed waveform in the true model and the modeled waveform using Model-0 Green's functions, for direct P waves. The time shift is used to account for unaccounted heterogeneities in the Green's function and is commonly used in earthquake source studies (Kikuchi and Kanamori, 1982).

If we take the Voigt notation to simplify tensors:

$11 \rightarrow 1, 22 \rightarrow 2, 33 \rightarrow 3, 12 \rightarrow 6, 13 \rightarrow 5, 23 \rightarrow 4,$

We have:

$$M_\alpha^J = m_{pq}^J$$
$$K_{k1}^0 = G_{k1,1}^0, K_{k2}^0 = G_{k2,2}^0, K_{k3}^0 = G_{k3,3}^0$$
$$K_{k4}^0 = G_{k2,3}^0 + G_{k3,2}^0$$
$$K_{k5}^0 = G_{k3,1}^0 + G_{k1,3}^0 \qquad (4)$$
$$K_{k6}^0 = G_{k1,2}^0 + G_{k2,1}^0$$
$$\alpha = 1,...,6;\ p,q,k = 1,2,3$$

By applying partial derivatives with respect to $M_\alpha^J$ and $\tau_\mathbf{r}^J$, respectively, we have:

$$\frac{\partial L(M_\beta^J, \tau_\mathbf{r}^J)}{\partial M_\alpha^J} = -2\sum_\mathbf{r}\sum_{k=1}^{3} \int \left[ u_k^J(\mathbf{r}, t+\tau_\mathbf{r}^J) K_{k\alpha}^0 - K_{k\alpha}^0 K_{k\beta}^0 M_\beta^J \right] dt = 0, \qquad (5)$$

and:

$$\frac{\partial L(M_\beta^J, \tau_\mathbf{r}^J)}{\partial \tau_\mathbf{r}^J} = 2\sum_{k=1}^{3} \int \left\{ \left[ u_k^J(\mathbf{r},t+\tau_\mathbf{r}^J) - K_{k\beta}^0 M_\beta^J \right] \frac{\partial u_k^J(\mathbf{r},t+\tau_\mathbf{r}^J)}{\partial \tau_\mathbf{r}^J} \right\} dt = 0. \qquad (6)$$

In our inversion, we first set $\tau_\mathbf{r} = 0$ for all seismograms and use equation (5) to determine the moment tensor components $M_\alpha^J$. We then put the inverted $M_\alpha^J$ into equation (6) to solve for $\tau_\mathbf{r}$ for each receiver. We repeat this procedure and iteratively determine both moment tensor $M_\alpha^J$ and the time shift, $\tau_\mathbf{r}$.

# 4  Numerical Results

We first present comparisons between the CLVD components in true moment tensors and the inverted moment tensors and for the 5 different models under three station coverage scenarios. The CLVD components were measured using the parameter $f_{CLVD}$ defined as:

$$f_{CLVD} = -\frac{\lambda_2}{\max(|\lambda_1|,|\lambda_3|)}, \tag{7}$$

where $\lambda_1 \geq \lambda_2 \geq \lambda_3$ are the three eigenvalues of the deviatoric moment tensor. The $f_{CLVD}$ value ranges from -0.5 to 0.5 with a value 0 for a pure DC source mechanism. The two extreme values $\pm 0.5$ are for pure CLVD source mechanisms. Positive $f_{CLVD}$ corresponds to extensional polarity for major dipole of the CLVD component. If we constrain the isotropic component to be zero (as what routine moment tensor inversion did), the percentage of CLVD component (Vavrycuk, 2005) in the moment tensor can be calculated as:

$$p_{CLVD} = 2|f_{CLVD}| \times 100\% \tag{8}$$

This percentage ranges from 0% to 100% with 0% represents pure DC and 100% represents pure CLVD source mechanism.

## 4.1 Case 1: Perfect station coverage

We will first analyze the situation with perfect station coverage where all 441 receivers were used for the least-squares inversion (Figure 3a). In this situation, we have full coverage for both up-going and down-going direct $P$ arrivals. Figure 4 contains the comparison between inverted $f_{CLVD}$ values and true $f_{CLVD}$ values for 24 earthquakes in five different models. We use $Err^{(J)}$ to denote the root mean squares (RMS, for the 24 earthquakes) error between inverted $f_{CLVD}$ values and true $f_{CLVD}$ values for Model-$J$ ($J = 1,2,3,4,5$). In this case, we have $Err^{(1)} \approx 0.008, Err^{(2)} \approx 0.023, Err^{(3)} \approx 0.036, Err^{(4)} \approx 0.008, Err^{(5)} \approx 0.007$. Notice that the true moment tensors are actually pure double couples for Model-1 and 4 because a shear dislocation source embedded in an isotropic homogeneous structure will not produce non-DC components. For Model-2 and 3, shear dislocations are embedded in a strong anisotropic medium and this will produce large $f_{CLVD}$ values for some earthquakes based on equation (1). The CLVD caused by existence of anisotropy varies with faulting geometry. For some special faulting geometry, the $f_{CLVD}$ value is zero even though the earthquake is embedded in anisotropy. The largest $f_{CLVD}$ value is about 0.126. For Model-5, the weakly anisotropic MOW contributes a small amount of CLVD component ($f_{CLVD} = 0.019$ on average) to the moment tensors. The inversion result shows that for perfect station coverage, the heterogeneous structure such as a subducting slab (Figure 2a) and the supra-slab weakly anisotropic structure (Figure 2d) has a very little effect on the $f_{CLVD}$ values of inverted moment tensors (Figure 4a, d). The inverted $f_{CLVD}$ values for Model-1 and 4 are close to zero (0.008 on average). For Model-2, 3, and 5 the CLVD components in the inverted moment tensors are actually caused by unaccounted source-side structures in the Green's function. In our study, the anisotropic structure is responsible for the apparent CLVD components in inverted moment tensors. Strong in-situ anisotropic patch could lead to large CLVD components in inverted moment tensors, while weakly anisotropic MOW alone could not produce large CLVD components observed in inverted moment tensors.

## 4.2 Case 2: Stations only in the lower hemisphere

In the real world, we may not have extensive station coverage for up-going direct arrivals. To make our station coverage consistent with the real world, we randomly selected 40 stations only from the lower hemisphere used for inversion (Figure 3b). Figure 5 contains the comparison between the inverted $f_{CLVD}$ values and true $f_{CLVD}$ values for this type of station coverage. For this poorer but more realistic station coverage, we notice that the $f_{CLVD}$ values in the inverted moment tensors have larger errors than those under the perfect station coverage. The RMS errors of inverted $f_{CLVD}$ values are $Err^{(1)} \approx 0.016, Err^{(2)} \approx 0.040, Err^{(3)} \approx 0.054, Err^{(4)} \approx$

$0.016, Err^{(5)} \approx 0.016$ for each model. Although the RMS errors doubled for Model-1, 4, and 5, our inverted moment tensors are still very close to the true mechanisms. This implies that with a reasonable station coverage, it is unlikely that apparent CLVD components of moment tensors are solely caused by source heterogeneous structure such as velocity heterogeneity in the subducting slab, supra-slab weakly anisotropic structure, or the MOW.

### 4.3 Case 3: Azimuth-biased lower-hemisphere station coverage

We then present comparisons between inverted moment tensors and true moment tensors for 5 different models with azimuth-biased lower-hemisphere station coverage.

We found that a very poor station coverage cause large errors in CLVD components of inverted moment tensors by investigating an azimuth-biased station distribution (Figure 3c). The RMS errors of $f_{CLVD}$ values for the 5 models are $Err^{(1)} \approx 0.038, Err^{(2)} \approx 0.058, Err^{(3)} \approx 0.097, Err^{(4)} \approx 0.035, Err^{(5)} \approx 0.033$, respectively. We observed that on average the RMS error of $f_{CLVD}$ in inverted moment tensors in Model-1 and 4 (Figure 6) is about 0.037 while the biggest one can almost reach 0.12 even though the true source mechanism is a pure double couple. The RMS errors for Model-2 and 3 also increases a lot. In Model-3, the strength of inverted $f_{CLVD}$ values for earthquake events 1 and 14 are 0.24 and 0.25, which is much larger than the true values 0.06 and 0.05. This indicates that very poor station coverage may introduce large errors to the CLVD components of the inverted moment tensors when the source is embedded in strong anisotropic medium. In Model-5, the RMS error doubles. The largest error in $f_{CLVD}$ values between inverted and true moment tensors is about 0.08.

## 4.4 Case 4: Same station distribution as Case 3 with the Green's function considering heterogeneous slab's effects

What if we include the information of the heterogeneous slab in our Green's function for Models with the existence of heterogeneous slab? To do so, we used the waveforms in Model-1 (Figure 2a) instead of those in Model-0 as the Green's functions in the MT inversion. We then perform the same inversion procedures to invert for moment tensors for Model-1, 3, 4, and 5 using the azimuth-biased station coverage. Figure 7 contains the comparison between the inverted CLVD components and true CLVD components. The RMS errors of CLVD components for five models are $Err^{(1)} \approx 0.001, Err^{(3)} \approx 0.066, Err^{(4)} \approx 0.006, Err^{(5)} \approx 0.027$. The error of CLVD components in Model-1 and 4 is decreased to almost zero. The inverted moment tensors are almost pure double-couple mechanisms which are even more precise than the case with perfect station coverage. However, for Model-3 and 5, although the error has been decreased by about 30% compared with the error in Case 3, the inverted $f_{CLVD}$ values still possesses large errors.

## 5 Discussion

In Model-1, we tested the effect of heterogeneous slab on the CLVD components of inverted moment tensors. Since the sources are embedded in locally isotropic homogeneous medium, the true source mechanism should be pure DC (zero CLVD). However, we found that for reasonable station coverage cases (Case 1 and Case 2), around 3.2% ($p_{CLVD}$) artificial CLVD component were introduced to the inverted moment tensors when effect of the slab was not considered in the kernel functions during moment tensor inversion. Even though artificial CLVD components can be introduced by the heterogeneous slab, its magnitude cannot match the prevalently observed large CLVD components (>10%). The observed apparent large CLVD components may represent the existence of other source properties such as strong anisotropic structure in the source region.

In Model-4, we mainly aim at testing whether supra-slab weakly anisotropic structure could cause apparent CLVD component in moment tensors. We decided this anisotropic patch dimension and anisotropy strength based on the shear-wave splitting studies where a typical travel time difference between fast and slow shear waves is about 1 second (Nowacki *et al.*, 2015). Our inversion results show that the supra-slab weakly anisotropic structure has nearly no effects on the CLVD components of inverted moment tensors compared with the heterogeneous slab because the inverted CLVDs in Model-4 are highly correlated with the ones in Model-1 for all four cases in this study (Figure 8).

Shear-wave splitting studies could tell the general travel time difference between fast wave and slow wave caused by the averaged anisotropy on the ray path. However, they could not resolve in which part on the raypath the anisotropic structure resides. Based on our numerical study, the supra-slab weakly anisotropic structure has nearly no effect on observed CLVD components while the in-situ anisotropic structure could cause significant CLVD components. The observation of large CLVD components thus indicates that the anisotropy observed using the shear-wave splitting method may be localized strong anisotropy residing in the source region and could be responsible for the occurrence of CLVD components in moment tensors.

In Case 4, we used the waveforms generated in Model-1 as new Green's functions for inverting moment tensors. The new Green's functions include information about the heterogeneous slab. For this case, the moment tensors could be even better inverted for an azimuth-biased station coverage for Model-1 and 4 compared with perfect station coverage. For Model-3 and Model-5 , the RMS error is decreased by about 30% compared with the RMS error using Green's function in an isotropic homogeneous medium. This suggests that it would be intriguing to consider the effect of the heterogeneous subducting slab in the kernel functions during routine inversion to improve the accuracy of the CLVD components.

We also test the effect of the weakly anisotropic MOW on the CLVD components of moment tensors in Model-5. We selected the MOW model with a similar dimension and velocity structures to the ones in previous studies (Kirby *et al.*, 1996; Jiang *et al.*, 2008). We found that the inverted CLVD component in all 4 cases in this study is reasonably close to the true values, and the MOW could only contribute a small amount of CLVD component to moment tensors. This indicates that MOW alone cannot produce the prevalently observed large CLVD components in the moment tensors of deep earthquakes. And there may exist other structures such as strong anisotropic fabric layers around deep earthquakes (Li *et al.*, 2018).

# 6 Conclusions

In this numerical study, we tested the effect of near source complexities including heterogeneous slab, supra-slab weakly anisotropic patch, intra-slab strongly anisotropic patch, and intra-slab weakly anisotropic wedge on the CLVD component of inverted moment tensors. We found that, under reasonable station coverages, the existence of the heterogeneous slab could cause artificial CLVD component in the inverted moment tensors. However, the magnitude of the artificial CLVD component cannot match the prevalently observed large CLVDs in routine moment tensor inversion. The supra-slab weakly anisotropic structure nearly has no effect on the CLVD components of inverted moment tensors. The intra-slab weakly anisotropic wedge (to imitate MOW) along also cannot produce CLVD with a similar magnitude to the ones obtained in routine moment tensor inversion. The strong intra-slab anisotropic patch could be a potential mechanism that could account for the observed significant CLVD components.

However, very poor station coverage (such as azimuth-biased coverage) could introduce large errors to the CLVD components of the inverted moment tensors. This error represents artificial CLVD components while the true source mechanism is pure DC or large deviation from true CLVD values while the source is deviated from pure-DC. This result indicates that sufficient station coverage (or azimuth coverage) is crucial in robust moment tensor inversion.

We also found that if we consider the effect of heterogeneous slab in the Green's function, the CLVD components would be much better resolved. This indicates that consideration of heterogeneous slab structure in kernel/or Green's functions of routine moment tensor inversion is highly recommended.

# 7 Data and Resources

This article uses synthetic dataset generated by 3D full-elastic finite-difference modeling. Data will be available upon request to the authors.

# 8 Acknowledgements

This work was supported by NSF EAR-1621878.

# 9 References

Aki, K., and P. G. Richards (2002), Quantitative seismology, University Science Books.

Bina, C. R., and B. J. Wood (1987), OLIVINE-SPINEL TRANSITIONS - EXPERIMENTAL AND THERMODYNAMIC CONSTRAINTS AND IMPLICATIONS FOR THE NATURE OF THE 400-KM SEISMIC DISCONTINUITY, J Geophys Res-Solid, **92**, 4853-4866.

Dziewonski, A. M., T. A. Chou, and J. H. Woodhouse (1981), Determination of earthquake source parameters from waveform data for studies of global and regional seismicity, Journal of Geophysical Research: Solid Earth, **86**, 2825-2852.

Ekström, G., M. Nettles, and A. M. Dziewonski (2012), The global CMT project 2004-2010: Centroid-moment tensors for 13,017 earthquakes, Phys. Earth Planet. Inter., **200**, 1-9.

Fang, X., Y. Zheng, and M. Fehler (2017), Fracture clustering effect on amplitude variation with offset and azimuth analyses, Geophysics, **82**, N13-N25.

Frohlich, C. (1990), Note concerning non-double-couple source components from slip along surfaces of revolution, Journal of Geophysical Research: Solid Earth (1978–2012), **95**, 6861-6866.

Furumura, T., B. L. N. Kennett, and S. Padhy (2016), Enhanced waveguide effect for deep-focus earthquakes in the subducting Pacific slab produced by a metastable olivine wedge, Journal of Geophysical Research: Solid Earth, **121**, 6779-6796.

Gilbert, F. (1971), Excitation of the normal modes of the earth by earthquake sources, Geophys J Roy Astr S, **22**, 223-226.

Green, H. W., 2nd, W. P. Chen, and M. R. Brudzinski (2010), Seismic evidence of negligible water carried below 400-km depth in subducting lithosphere, Nature, **467**, 828-831.

Henry, C., J. H. Woodhouse, and S. Das (2002), Stability of earthquake moment tensor inversions: effect of the double-couple constraint, Tectonophysics, **356**, 115-124.

Jiang, G., D. Zhao, and G. Zhang (2008), Seismic evidence for a metastable olivine wedge in the subducting Pacific slab under Japan Sea, Earth and Planetary Science Letters, **270**, 300-307.

Johnston, D. E., and C. A. Langston (1984), The effect of assumed source structure on inversion of earthquake source parameters - the Eastern Hispaniola earthquake of 14 September 1981, B Seismol Soc Am, **74**, 2115-2134.

Julian, B. R., A. D. Miller, and G. R. Foulger (1998), Non-double-couple earthquakes 1. Theory, Reviews of Geophysics, **36**, 525-549.

Kawakatsu, H., and S. Yoshioka (2011), Metastable olivine wedge and deep dry cold slab beneath southwest Japan, Earth and Planetary Science Letters, **303**, 1-10.

Kawasaki, I., and T. Tanimoto (1981), Radiation patterns of body waves due to the seismic dislocation occurring in an anisotropic source medium, B Seismol Soc Am, **71**, 37-50.

Kikuchi, M., and H. Kanamori (1982), Inversion of complex body waves, Bulletin of the Seismological Society of America, **72**, 491-506.

Kirby, S. H., S. Stein, E. A. Okal, and D. C. Rubie (1996), Metastable mantle phase transformations and deep earthquakes in subducting oceanic lithosphere, Reviews of Geophysics, **34**, 261-306.

Knopoff, L., and M. J. Randall (1970), The compensated linear-vector dipole: A possible mechanism for deep earthquakes, Journal of Geophysical Research, **75**, 4957-4963.

Kubo, A., E. Fukuyama, H. Kawai, and K. i. Nonomura (2002), NIED seismic moment tensor catalogue for regional earthquakes around Japan: quality test and application, Tectonophysics, **356**, 23-48.

Kuge, K., and H. Kawakatsu (1993), Significance of non-double couple components of deep and intermediate-depth earthquakes: implications from moment tensor inversions of long-period seismic waves, Physics of the Earth and Planetary Interiors, **75**, 243-266.

Li, J., Y. Zheng, L. Thomsen, T. J. Lapen, and X. Fang (2018), Deep earthquakes in subducting slabs hosted in highly anisotropic rock fabric, Nature Geoscience, **11**, 696-700.

Nowacki, A., J. M. Kendall, J. Wookey, and A. Pemberton (2015), Mid-mantle anisotropy in subduction zones and deep water transport, Geochemistry, Geophysics, Geosystems, **16**, 764-784.

Randall, M. J., and L. Knopoff (1970), The mechanism at the focus of deep earthquakes, Journal of Geophysical Research, **75**, 4965-4976.

Satake, K. (1985), Effects of station coverage on moment tensor inversion, B Seismol Soc Am, **75**, 1657-1667.

Tada, T., and K. Shimazaki (1994), How much does a high-velocity slab contribute to the apparent non-double-couple components in deep-focus earthquakes, B Seismol Soc Am, **84**, 1272-1278.

Vavrycuk, V. (2004), Inversion for anisotropy from non-double-couple components of moment tensors, J Geophys Res-Sol Ea, **109**.

Vavrycuk, V. (2005), Focal mechanisms in anisotropic media, Geophys J Int, **161**, 334-346.

Woodhouse, J. H. (1981), The excitation of long period seismic waves by a source spanning a structural discontinuity, Geophysical Research Letters, **8**, 1129-1131.

# 10 Full mailing address

Jiaxuan Li: Science and Research Building 1, Department of Earth and Atmospheric Sciences, University of Houston, Houston, TX, USA.

Yingcai Zheng: Science and Research Building 1, Department of Earth and Atmospheric Sciences, University of Houston, Houston, TX, USA.

Xinding Fang: Department of Ocean Science and Engineering, Southern University of Science and Technology of China, No 1088, Xueyuan Rd., Nanshan District, Shenzhen, Guangdong 518055, China

# 11 List of Figure Captions

Figure 1. (a) The source and receiver geometry in FD forward modeling. The red star represents the source. The blue circles represent the receivers. The 441 receivers are evenly distributed on a sphere with a radius of 500km centered at the source. The background $V_P = 8 km/s$ and $V_S$=4.5 km/s. (b) P and S velocities in the heterogeneous slab for model-1, 2, 4, and 5 (Figure 2a, b, d, e). The highest velocities in the slab are $V_P = 8.4 km/s$ and $V_S$=4.7 km/s, which represent about 5% high-velocity anomaly in the coolest part of the slab.

Figure 2. Five different models (only showing the P wave velocities) with a heterogeneous slab or an anisotropic structure. The dipping angle of the slab is 40°. The red star represents the source location. The size of heterogeneous slab is $100km \times 600km \times 600km$. The anisotropic structure is chosen to be tilted transversely isotropic (TTI) medium, which is described by the $V_P$ and $V_S$ (same as background $V_P$ and $V_S$) along the TTI symmetry axis, and three Thomsen parameters $\epsilon, \gamma, \delta$. In Model-2, 3, and 4, the TTI symmetry axis is perpendicular to the slab interface. In Model-5, the TTI symmetry axis is parallel to the slab interface. For Model-2 and 3, the dimension of the anisotropic patch is $30km \times 30km \times 30km$ and the three Thomsen parameters are $\varepsilon = 20\%, \gamma = 20\%, \delta = 0\%$, which represents a strong in-situ anisotropic structure. For model-4, the anisotropic patch is $150km \times 150km \times 20km$ and the $\varepsilon = 0\%, \gamma = 6\%, \delta = 0\%$ which represents a weak outside slab anisotropic structure. For Model-5, the $\varepsilon = 0\%, \gamma = 5\%, \delta = 0\%$, which represents a weakly anisotropic MOW inside the slab. (a) Model-1 with only heterogeneous slab. (b) Model-2 with only strong anisotropic patch in the source region. (c) Model-3 with both heterogeneous slab and strong anisotropic patch in the source region. (d) Model-4 with heterogeneous slab and weakly anisotropic patch outside the slab. (e)

Model-5 with heterogeneous slab and weakly anisotropic MOW inside the slab. (f) 2D velocity profile of Model-5. The distance of three tips of MOW A, B, and C to the slab upper interface is 20km, 0km, and 40km.

Figure3. Stereographic projection of station distributions for three cases. The arrow represents North. The red triangles represent receivers on the upper hemisphere and blue triangles represent receivers on the lower hemisphere. (a) Perfect station coverage including all 441 stations. (b) 40 randomly selected stations in the lower hemisphere are randomly selected. (c) Azimuth-biased station coverage with 40 stations in the lower hemisphere in the southern part selected.

Figure 4. Perfect station coverage - comparison of $f_{CLVD}$ values between inverted moment tensors and true moment tensors for the 24 earthquake events, for the five models. (a) Model-1: The inverted $f_{CLVD}$ values are very close to the true ones, which are essentially zeros. (b) Model-2: The inverted $f_{CLVD}$ values are close to the true ones. Some of the $f_{CLVD}$ values are not zero because they are caused by faulting in the anisotropic structure in the source region. (c) Model-3: The inverted $f_{CLVD}$ values are still close to the true ones. The error is a little bit larger than the results in Model-2. (d) Model-4: We have similar results to the ones in Model-1. The supra-slab weakly-anisotropic patch has nearly no effect. (e) Model-5: Inverted $f_{CLVD}$ values are very close to the true ones.

Figure 5. Lower-hemisphere station coverage - comparison of $f_{CLVD}$ values between true and inverted moment tensors for the 24 earthquake events, for the five models. The moment tensors are inverted using 40 randomly selected stations only in lower hemisphere. For this station coverage, the inverted $f_{CLVD}$ values have larger errors for all 4 Models. But the results are still reasonable. (a-e) Comparison for Model-1~5.

Figure 6. Azimuth-biased lower-hemisphere coverage – comparison of $f_{CLVD}$ values between true moment tensors and inverted moment tensors for the 24 earthquakes, for the five models. The moment tensors are inverted using 40 randomly selected azimuth-biased stations. This poor station distribution contributes large errors to the inverted $f_{CLVD}$ values. (a-e) Comparison for Model-1~5.

Figure 7. Azimuth-biased lower-hemisphere coverage (same coverage as the one in Figure 6). We used waveforms in Model-1 as the new Green's functions. The new Green's function considers the effect of the isotropic heterogeneous slab. Comparisons of $f_{CLVD}$ values between the true moment tensors and the inverted moment tensors using the new Green's functions for the 24 earthquakes, for the four models: (a) Comparison for Model-1. (b) Comparison for Model-3. (c) Comparison for Model-4. (d) Comparison for Model-5.

Figure 8. Comparison between the inverted CLVD component in Model-1 and Model-4 for 4 cases. The inverted $f_{CLVD}$ values in Model-4 are highly correlated with the ones in Model-1.

# 12 Figures

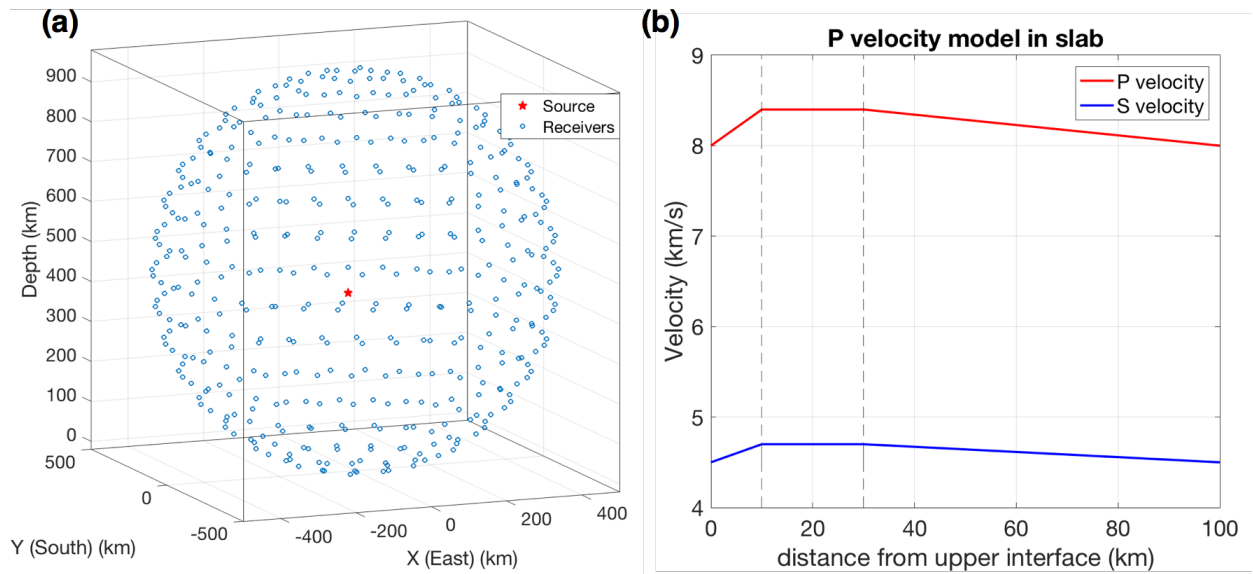

Figure 1. (a) The source and receiver geometry in FD forward modeling. The red star represents the source. The blue circles represent the receivers. The 441 receivers are evenly distributed on a sphere with a radius of 500km centered at the source. The background $V_P = 8km/s$ and $V_S$=4.5 km/s. (b) P and S velocities in the heterogeneous slab for model-1, 2, 4, and 5 (Figure 2a, b, d, e). The highest velocities in the slab are $V_P = 8.4km/s$ and $V_S$=4.7 km/s, which represent about 5% high-velocity anomaly in the coolest part of the slab.

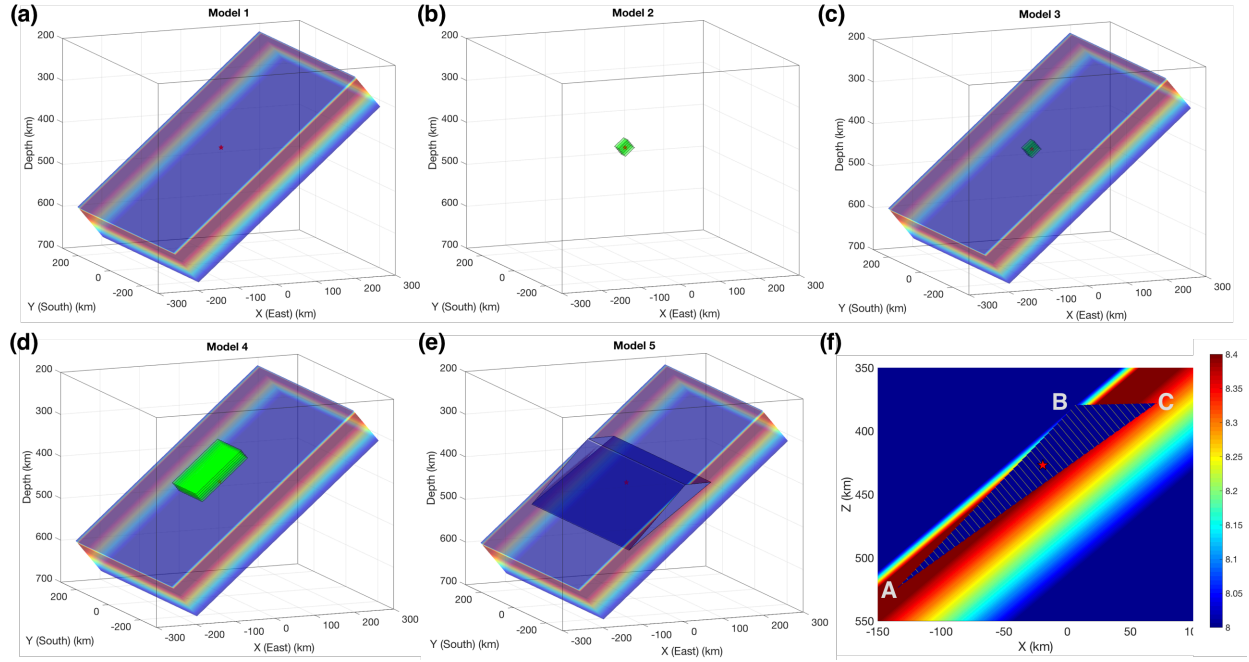

Figure 2 Five different models (only showing the P wave velocities) with a heterogeneous slab or an anisotropic structure. The dipping angle of the slab is 40°. The red star represents the source location. The size of heterogeneous slab is $100km \times 600km \times 600km$. The anisotropic structure is chosen to be tilted transversely isotropic (TTI) medium, which is described by the $V_P$ and $V_S$ (same as background $V_P$ and $V_S$) along the TTI symmetry axis, and three Thomsen parameters $\epsilon, \gamma, \delta$. In Model-2, 3, and 4, the TTI symmetry axis is perpendicular to the slab interface. In Model-5, the TTI symmetry axis is parallel to the slab interface. For Model-2 and 3, the dimension of the anisotropic patch is $30km \times 30km \times 30km$ and the three Thomsen parameters are $\varepsilon = 20\%, \gamma = 20\%, \delta = 0\%$, which represents a strong in-situ anisotropic structure. For model-4, the anisotropic patch is $150km \times 150km \times 20km$ and the $\varepsilon = 0\%, \gamma = 6\%, \delta = 0\%$ which represents a weak outside slab anisotropic structure. For Model-5, the $\varepsilon = 0\%, \gamma = 5\%, \delta = 0\%$, which represents a weakly anisotropic MOW inside the slab. (a) Model-1 with only heterogeneous slab. (b) Model-2 with only strong anisotropic patch in the source region. (c) Model-3 with both heterogeneous slab and strong anisotropic patch in the source region. (d) Model-4 with heterogeneous slab and weakly anisotropic patch outside the slab. (e) Model-5 with heterogeneous slab and weakly anisotropic MOW inside the slab. (f) 2D velocity profile of Model-5. The distance of three tips of MOW A, B, and C to the slab upper interface is 20km, 0km, and 40km.

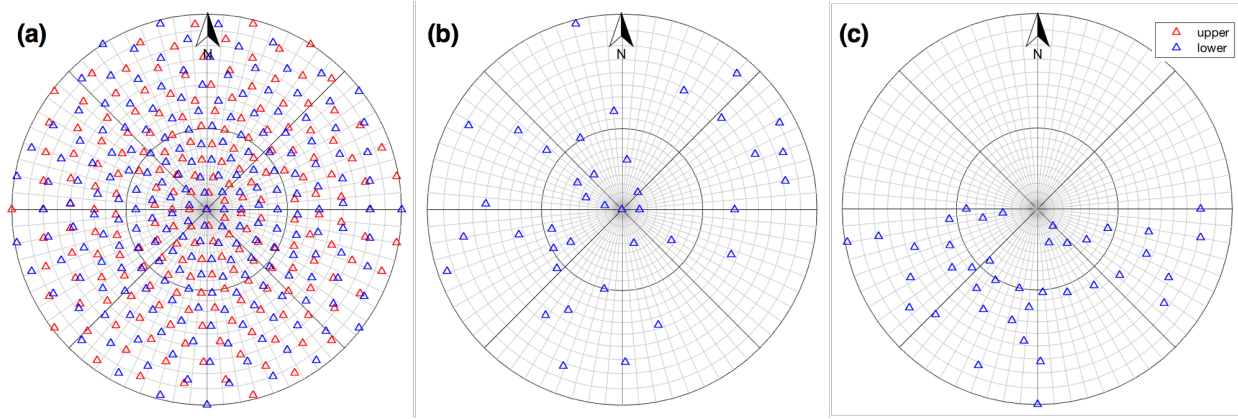

Figure 3. Stereographic projection of station distributions for three cases. The arrow represents North. The red triangles represent receivers on the upper hemisphere and blue triangles represent receivers on the lower hemisphere. (a) Perfect station coverage including all 441 stations. (b) 40 randomly selected stations in the lower hemisphere are randomly selected. (c) Azimuth-biased station coverage with 40 stations in the lower hemisphere in the southern part selected.

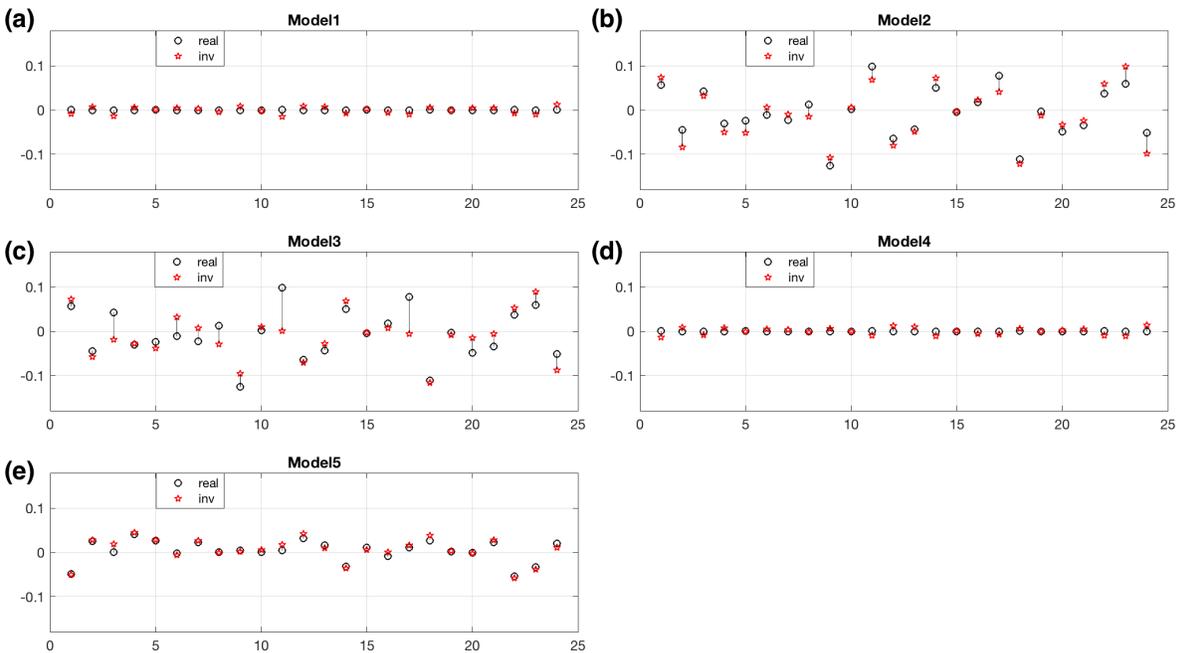

Figure 4. Perfect station coverage - comparison of $f_{CLVD}$ values between inverted moment tensors and true moment tensors for the 24 earthquake events, for the five models. (a) Model-1: The inverted $f_{CLVD}$ values are very close to the true ones, which are essentially zeros. (b) Model-2: The inverted $f_{CLVD}$ values are close to the true ones. Some of the $f_{CLVD}$ values are not zero because they are caused by faulting in the anisotropic structure in the source region. (c) Model-3: The inverted $f_{CLVD}$ values are still close to the true ones. The error is a little bit larger than the results in Model-2. (d) Model-4: We have similar results to the ones in Model-1. The supra-slab weakly-anisotropic patch has nearly no effect. (e) Model-5: Inverted $f_{CLVD}$ values are very close to the true ones.

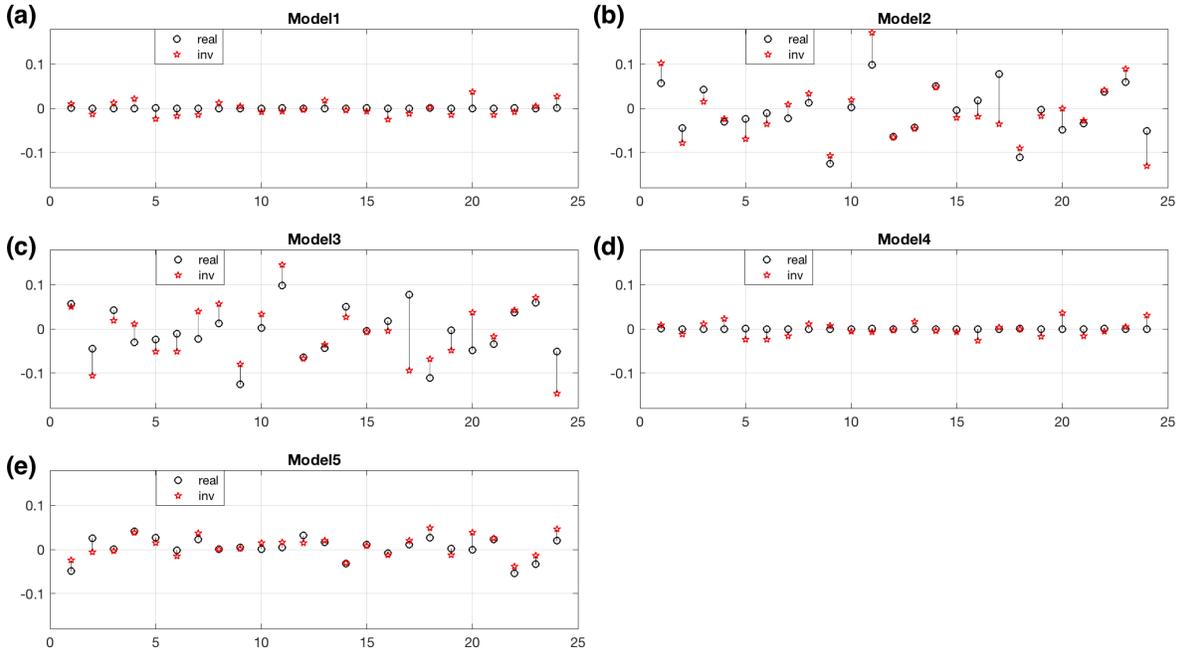

Figure 5. Lower-hemisphere station coverage - comparison of $f_{CLVD}$ values between true and inverted moment tensors for the 24 earthquake events, for the five models. The moment tensors are inverted using 40 randomly selected stations only in lower hemisphere. For this station coverage, the inverted $f_{CLVD}$ values have larger errors for all 4 Models. But the results are still reasonable. (a-e) Comparison for Model-1~5.

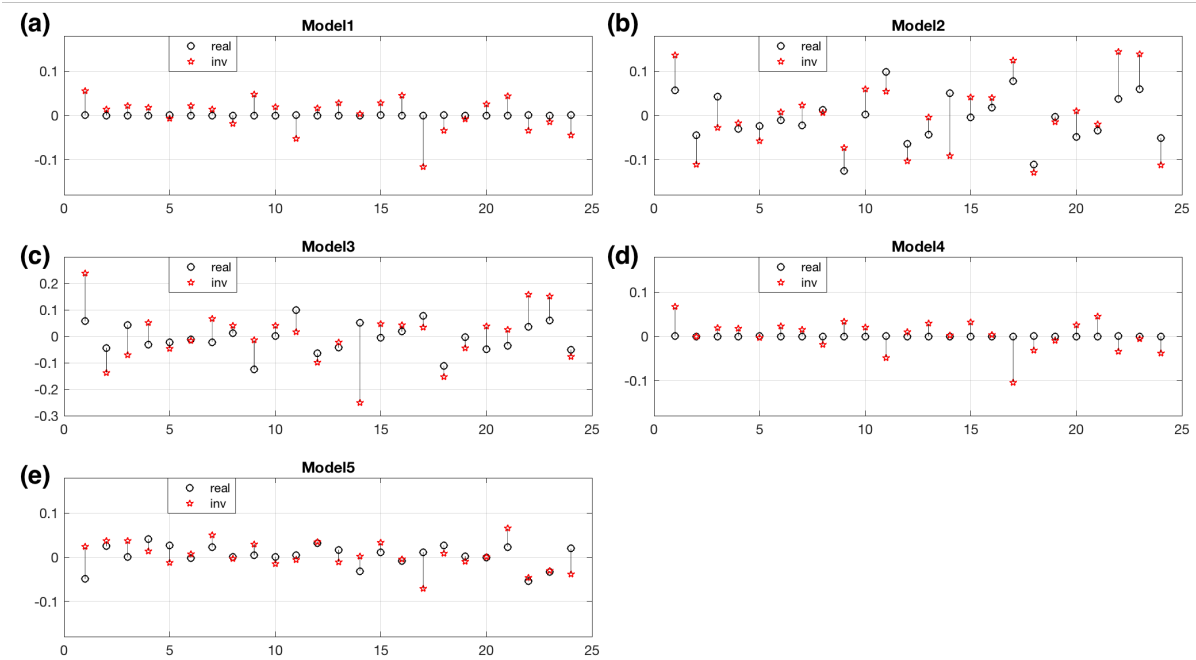

Figure 6. Azimuth-biased lower-hemisphere coverage – comparison of $f_{CLVD}$ values between true moment tensors and inverted moment tensors for the 24 earthquakes, for the five models. The moment tensors are inverted using 40 randomly selected azimuth-biased stations. This poor station distribution contributes large errors to the inverted $f_{CLVD}$ values. (a-e) Comparison for Model-1~5.

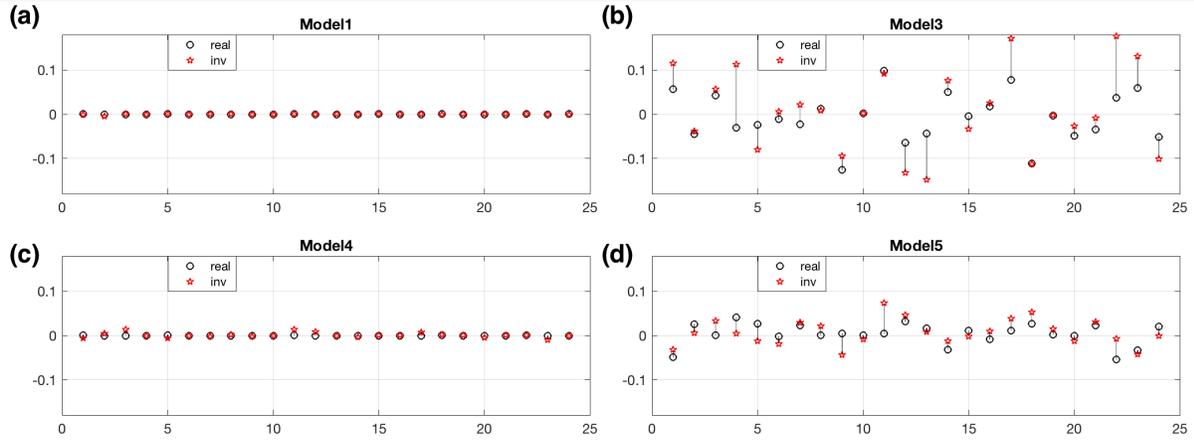

Figure 7. Azimuth-biased lower-hemisphere coverage (same coverage as the one in Figure 6). We used waveforms in Model-1 as the new Green's functions. The new Green's function considers the effect of the isotropic heterogeneous slab. Comparisons of $f_{CLVD}$ values between the true moment tensors and the inverted moment tensors using the new Green's functions for the 24 earthquakes, for the four models: (a) Comparison for Model-1. (b) Comparison for Model-3. (c) Comparison for Model-4. (d) Comparison for Model-5.

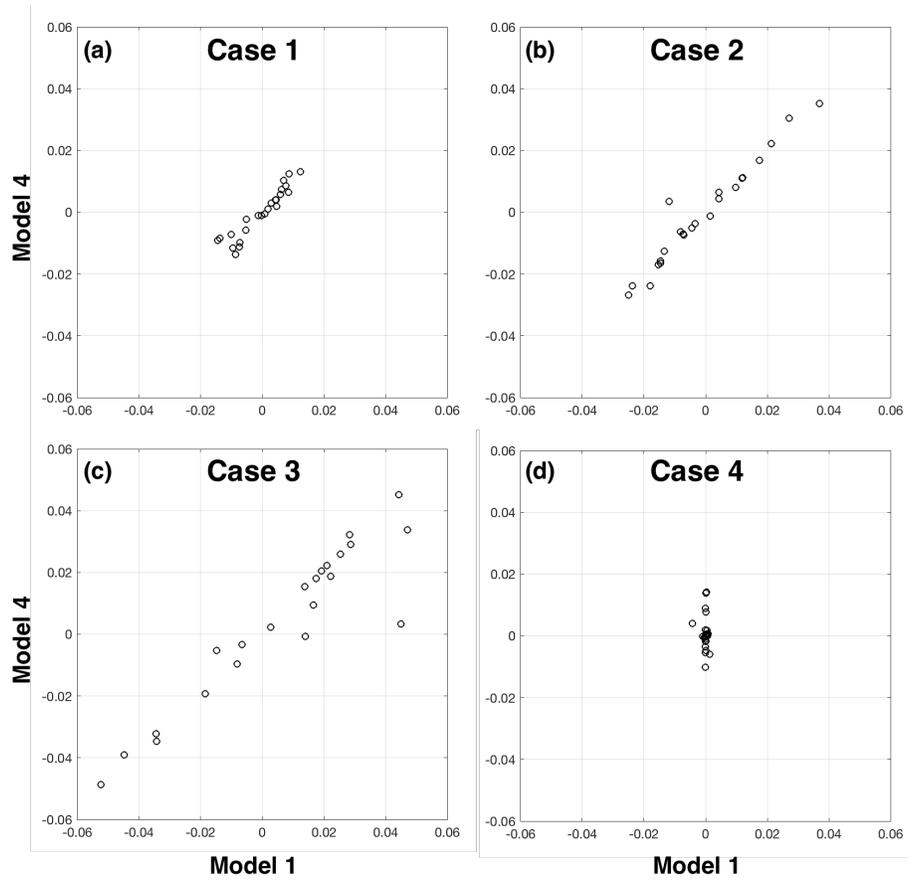

Figure 8. Comparison between the inverted CLVD component in Model-1 and Model-4 for 4 cases. The inverted $f_{CLVD}$ values in Model-4 are highly correlated with the ones in Model-1.